# Image processing of a spectrogram produced by Spectrometer Airglow Temperature Imager

*Atanas Marinov Atanassov*

*Solar-Terrestrial Influences Laboratory, Bulgarian Academy of Sciences, Stara Zagora Division, Bulgaria,*
*At_M_Atanassov@yahoo.comy*

**Introduction**

The Spectrometer Airglow Temperature Imager (SATI) is a ground-based spatially scanning Fabry-Perot interferometer [1].

Basically, the stages of the SATI image processing and analysis are analogous to those already considered in earlier works [2,34,5]. However, the tools and approach for their implementation, described in the present paper, are substantially different.

The basic ideas for the instrument and data processing were developed about two decades ago at the York University - Canada [2, 3, 4].

A new version of SATI [6] was developed at the Stara Zagora Division of the Solar-Terrestrial Influences Laboratory.

The objective of this paper is to present a part of the overall work, connected with the updating of the SATI data processing algorithms.

**Overview of instrument operation**

The operation of the SATI-3 instrument is automatic and PC-controlled. The measurements start after astronomical twilight in the beginning of the night and finish before its beginning in the morning, before sunrise and when the Moon is under the horizon. Each spectrogram is obtained with exposure time of 120 s. As a result of the measurement sequence, two series of images are accumulated, from which data are extracted.

– $I_{m,n}(t_{im})$, im=1,...,P- spectrogram images
– $D_{m,n}(t_d)$, d=1,...,Q- dark current images

After each measurement sequence, depending on its duration, spectrograms with P number and Q dark images are available. The (m,n) indices determine the location of the pixels in the bi-dimensional image. The time step, when dark current images are obtained, is almost constant. The images, containing spectrograms are also formed with an almost constant time step. After each sequence of eight observation images a dark current image is taken.

The dark current images are influenced, although slightly, by the CCD temperature, which varies. The stabilization of the interference filter temperature is of major importance for maintaining its parameters (μ- refractive index, $\lambda_0$- maximum transmission wavelength) within certain limits and for the correct spectral lines identification.

**Preliminary image processing**

*Correction of high-value pixels*

The preliminary processing of image rows $I_{m,n}(t_i)$ and $D_{m,n}(t_d)$ is associated with the presence of instrumental noise as well as with a comparatively small number of CCD pixels, whose increased sensibility differs from that of the adjacent ones. In addition, due to the presence of high-energy charged particles, which fall on separate CCD pixels during exposition, the ADUs (Analog Digital Unit) values of the latter have a random character and, what is more – their values are increased.

Figure 1,a shows an example of a spectrogram image. The Dark current image looks like the corner of a spectrogram image – without any particular structures and with noise. Bright points are visible resulting from electric charged particles, which have passed through the detector. The correction of the values of the so-called "hot pixels" and charged particles-hit pixels is made after their identification. Figure 2 shows the histograms of the pixels distribution by ADU values of one dark current image (a) and one of a spectrogram (b). It is seen that the main part of the dark current histogram is a symmetric Gaussian type, however, the tail to the right shows the presence of pixels with increased values – namely those which are searched for. The histogram of the spectrogram image is more complex due to the image structures – the low "hunch" refers to the area, outside the histogram, while the high "hunch" is for the area, containing the spectrogram. Here again there is a tail, containing the high-value pixels.

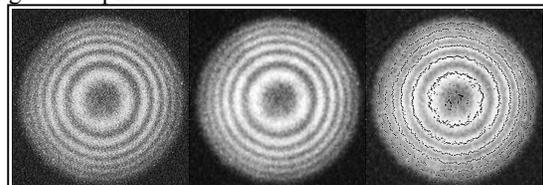

**Figure 1.** a) Original image; b) Removed high-energy particles vestige; and c) Ring-like

The histogram analysis provides the opportunity to find the pixels, whose ADU values are increased. This is performed

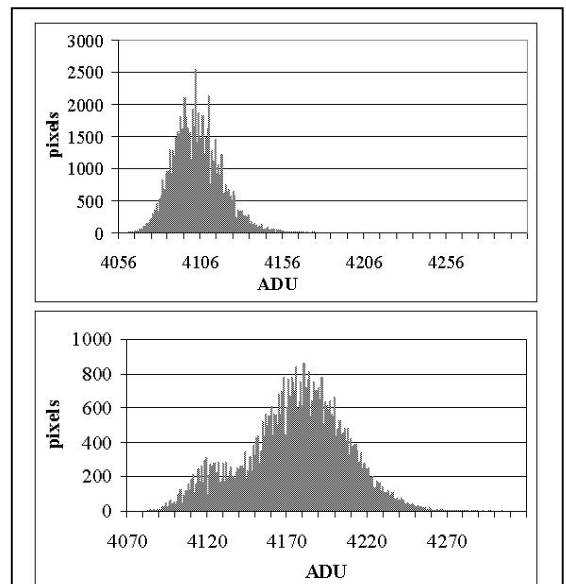

**Figure 2.** a) Histogram of dark image; b) Histograms of spectrogram image

on the basis of flexible criteria of statistical character. Figure

1,b shows an image after the application of the respective processing procedure.

The corrected values for dark image pixels could be expressed as:

$$D_{m,n}(t_d) = \begin{cases} (D_{m-1,n} + D_{m+1,n} + D_{m,n-1} + D_{m,n+1})/4 \\ \frac{1}{M.N}\sum_{m,n}^{M,N} D_{m,n} \quad \text{- mean value} \\ H_{max} \quad \text{- maximum of histogramm} \end{cases}$$

The processing of the spectrogram image is divided into two parts: 1) analysis of the image corners; and 2) analysis of the central part of the image. These will be discussed in turn.

*Dark image correction*

The histogram analysis of the dark image allows to determine the most probable ADU value for each image, hence – the nocturnal cycle of the most probable value. The amplitude of this course varies sometimes within the range of dozens of units while, in other cases – it is relatively constant. The ADU time course analysis for each pixel displays individual peculiarities with certain boundaries, which are not negligible. This is shown after a suitable filtration of the high-frequency noise:

$$D'_{m,n}(t_d) = \Phi_w^t(D_{m,n}(t_d)) = \Phi_w^t(D^0_{m,n}(t_d) + \Psi_{m,n}(t_d)),$$

$$D'_{m,n}(t_d) \approx D^0_{m,n}(t_d)$$

Figure 3 presents a comparison of the ADU values drift for a part of an image. Both the amplitude and the absolute values, as well as the course of each of the adjacent pixels shown (arbitrary selected fragment) are different. After determining the so-called Dark current image drift we can apply the following correction approach referring to the time $t_{im}$: $t_d \leq t_{im} \leq t_{d+1}$ of the spectrogram image formation:

$$D^*_{m,n}(t_{im}) = D_{m,n}(t_d) + (t_{im} - t_d)/(t_{d+1} - t_d)[D_{m,n}(t_{d+1}) - D_{m,n}(t_d)]$$

$D^*_{m,n}(t_{im})$ is the Dark current image, calculated on the basis of linear interpolation and reflects better both the specific character of the individual pixels and the time changes. The comparison shows a certain difference in determining the parameters of the interference filter ($\mu$-refraction index, $\lambda_0$- wavelength of maximum permeability) on the basis of the presented method for Dark current image correction with regards to the averaging of all images.

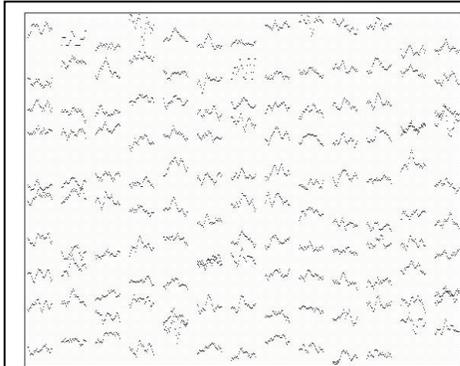

**Figure 3.** Individual pixel drift for a dark image fragment.

After subtracting the calculated dark current image value, we obtain the measured signal:

$$S_{m,n}(t_{im}) = r.[I_{m,n}(t_{im}) - D^*_{m,n}(t_{im})]$$

where r is the correction factor [5].

*Signal image filtering*

The image of the measured signal can be presented as a sum of the real signal $S^0_{m,n}(t_{im})$, received by the instrument and random noise $\Psi_{m,n}(t_{im})$. Then we can apply the following filtration:

$$S'_{m,n}(t_{im}) = \Phi_w^s(S_{m,n;w}(t_{im})) = \Phi_w^s(S_{m,n;w}(t_{im}) + \Psi_{m,n;w}(t_{im}))$$

where $\Phi_w^s$ is a 2-D filter and w is the respective window. The filtration of the measured signal is possible both in the pixel space and in time (corresponding to the dark-current image) in order to eliminate the high-frequency component in the signal. Considering the problem to be solved, a space filtration of the signal field is applied. The use of a 2-D running average value is an efficient high-frequency filter, applied for image processing [7]. Regarding the connection of the amplitude-frequency characteristics of this filter to the window length, on the one side, and the character of the structures in the signal 2-D field of application, on the other hand, a double filtering is selected with a window of 3x3 pixels. A special algorithm has been designed for simultaneous double filtration, thus minimizing the processing time.

**Determination of the interference filter parameters**

It is possible to extract information from each spectrogram, associated – on the one hand with the maximums' location and, on the other - with frequency energy distribution in the respective spectral part. The filter parameters, which are dependent on its temperature, are defined on the basis of the maxima position (interference rings radii), by solving the reverse problem. The following relation is used:

$$\frac{r_i^2}{r_i^2 + f^2} = \mu^2\left(1 - \left(\lambda_i/\lambda_0\right)^2\right), \quad i = \overline{1,6} \quad (1)$$

where $r_i$ is the radius of the $i^{th}$ ring, f is the focal distance of the objective and $\lambda_i$ is the respective wavelength. The unknown quantities in (6) are determined on the basis of a linear regression.

The filter parameters ($\mu, \lambda_0$) have been determined using Eq. 1. Three to six interference maxima/rings are used (the minima are not used). The deviation of the refraction index value, determined by the above-described spectrogram analysis, from the measured laboratory values ($\mu$=2.1551; 30˚C) in some cases can be of the order of a few thousands. The values of $\lambda_0$ vary in the 5-th sign, respectively (see Table. 2).

The following approach is applied here to determine the respective radii. First, an initial approximation of the interference rings centre is selected $O^0(m^0,n^0)$ and on the basis of the image radial sections (Fig. 4), the locations of the respective maxima are specified.

$$R_{k,\varphi} = S'_{k.\cos(\varphi),k.\sin(\varphi)}; \quad k = 1 \div 128;$$

$$\varphi = 0 \div 360°; \quad \Delta\varphi = 1°$$

The problem for maximum searching cannot always be solved. An approach has been developed for searching and recognition of maxima, which are not sufficiently distinguished in the presence of residual noise. Every interference maximum possesses own individuality - the central ones are better expressed, while the outer ones sometimes and in some parts of the image might appear smeared and not very clear. The search is not on the basis of comparing values of signals, by forming interference spectra by sectors, - it is done on the basis of space searching of conditions, based on pattern recognition. Every pattern $P_k(\{b_i\}i=1,L;L)$ is an arranged binary multitude $\{b_1, b_2, \ldots ,b_L\}$, as each element describes a relation between two adjacent points (increase-decrease) from the section. Since the search procedure is not always productive, it is necessary to reduce the criteria and more or less recognize an interference maximum in the space of relations with a suitable pattern. This approach provides the opportunity to increase the number of the interference maxima found, especially for the external rings. This is of major importance for the interference image with weaker contrast obtained with not very strong useful signal.

The pattern in the first row in Table 1 refers to cases when the data noise has not resulted in any obstruction of the maximum profile. The search of maxima is performed within the section frames, without any prior information for their position. The algorithms are applicable both to the specific spectrograms of $O_2$ and in similar cases, regardless of the measurement filter.

Table 1.

| patterns | elements | | | | | |
|---|---|---|---|---|---|---|
|  | 0 | 0 | 0 | 1 | 1 | 1 |
| 1-st | 0 | 0 | 0 | 0 | 0 | 1 |
| 2-st | 0 | 0 | 0 | 0 | 1 | 0 |
| 3-st | 0 | 0 | 1 | 0 | 1 | 0 |
| 4-th | 0 | 0 | 0 | 1 | 0 | 0 |
| 5-th | 0 | 0 | 0 | 1 | 1 | 0 |

After determining the coordinates of the interference maxima within the frames of the radial sections it is necessary to allocate them according to the interference rings. This is required because the maxima of all rings are not always found. In such cases the problem is solved on the basis of the analysis of the histogram for distribution of all determined maxima. The points for each ring with their coordinates are taken around the respective maxima in the histogram (Fig. 1c).

Having enough available points for each ring, the Least Square Fitting (LSF) can be applied to determine the centres and their radii. A multitude of points $\{(x_k,y_k)\}_{k, l=1,N}$ for each $k^{-th}$ ring is obtained, which should satisfy the condition to lie as close as possible to a circle with radius $r$ and centre coordinates $(a,b)$, that is $(x-a)^2 + (y-b)^2 = r^2$ have to be fulfilled. Since the determined points do not lie exactly on the circle, in order to define that circle, it is necessary to minimize the following functional:

$$E(a,b,r) = \sum_{l=1}^{N}(L_l - r)^2,$$
$$L_l = \sqrt{(x_l - a)^2 + (y_l - b)^2} \qquad (2)$$

The functional $E(a,b,r)$ is minimized by an approach, which does not lead to solving a non-linear LSF problem. An iteration procedure is applied [8] (Eberly, 2000), which is a convergence and yields very good results.

$$a = \bar{x} + \bar{L}\bar{L}_a \ , \ b = \bar{y} + \bar{L}\bar{L}_b, \qquad (3a)$$

$$\bar{x} = \frac{1}{N}\sum_{l=1}^{N}x_l \ , \ \bar{y} = \frac{1}{N}\sum_{l=1}^{N}y_l, \qquad (3b)$$

$$\bar{L} = \frac{1}{N}\sum_{l=1}^{N}L_l \ , \ \bar{L}_a = \frac{1}{N}\sum_{l=1}^{N}\frac{a - x_l}{L_l},$$

$$\bar{L}_b = \frac{1}{N}\sum_{l=1}^{N}\frac{b - y_l}{L_l} \qquad (3c)$$

Before starting the iteration process, as an initial approximation of the centre coordinates it is assumed that $a_0 = \bar{x}, b_0 = \bar{y}$. After completing the iteration process and defining the centre coordinates its radius is determined as:

$$r = \frac{1}{N}\sum_{l=1}^{N}L_l \qquad (4)$$

The entire process of defining the ring parameters is applied iteratively for their improvement. This process requires a good initial approximation of the image centre and it is quickly convergent. The proof for the process convergence is illustrated numerically. Figure 5 shows histograms of three sequential approximations. In the first histogram the three internal rings are well separated, in contrast to the three external rings. The next approximation of the spectrogram centre is selected to be one of the three internal rings. It can be seen that the third approximation does not improve substantially the maxima separation; the process for determination of the image centre and the rings radii is considered completed. The next step is the regression problem to determine the filter parameters $(\mu,\lambda_0)$. Tables 2 and 3 show the mean values of the interference rings radii and the image centre coordinates, corresponding to each of those rings, with the respective dispersion for an entire observation sequence. The mean values of the filter parameters are also shown. Figure 6 presents the nocturnal course of the filter parameters.

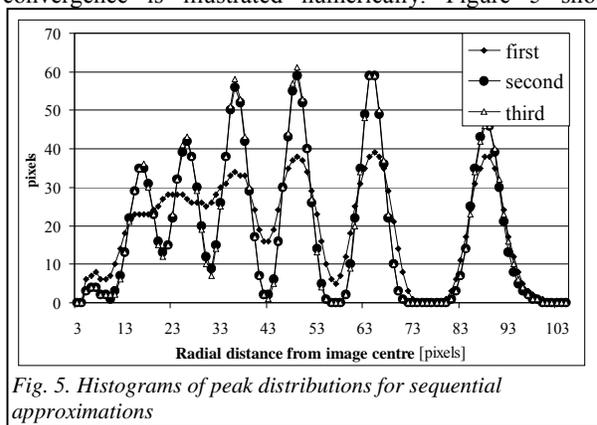

*Fig. 5. Histograms of peak distributions for sequential approximations*

**Conclusion**

The suggested approach is an attempt to look on the processing of the spectrograms, obtained by the SATI instrument in a different non-conventional way. A comparison will be necessary between the offered processing approaches with the already known ones.

A new approach for sector temperature determination is developed, which will be presented later on. It offers enhanced flexibility in regards to the image processing and a

new approach for mean temperature calculation without spectrum averaging

REFERENCES
[1] Shepherd, G.G. Spectral Imaging of the Atmosphere. Academic Press, International Geophysics Series, v. 82, 2002.
[2] Wiens R. H., S.-P. Zhang, R. N. Peterson and G. G. Shepherd, MORTI: a mesopause oxygen rotational temperature imager, Planet Space Sci., 39, 1363, 1991.
[3] Wiens R. H., A. Moise, S. Brown, S. Sargoytchev, R. N. Peterson, G. G. Shepherd, M. J. López-González, J. J. Lopez-Moreno and R. Rodrigo, SATI: a spectral airglow temperature imager, Adv. Space Res., 19, 677-680, 1997
[4] Zhang S. P., Gravity Waves from $O_2$ Airglow, PhD Thesis, York University, 1991.
[5] Cho, Y-M., Studies of the Polar MLT Region Using SATI Airglow Measurements, PhD Thesis, York University, 2006.
[6] N. Petkov, et al., SATI-3SZ - a Spectral Airglow Temperature Imager in Stara Zagora Station: Possibilities and First Results, Proceedings of conference "Fundamental Space Research", Sunny Beach, Bulgaria, 21-28 Sep 2008
[7] Glasbey C. A., R. Jones, Fast computation of moving average and in octagonal windows. Pattern Recognition Letters, 18, 555-565, 1997
[8] Eberly, D., "3D Game Engine Design", Morgan Kaufmann Publishers, San Francisco, CA 476-478, September 2000

Table 2

|  | $R_1$ | $R_2$ | $R_3$ | $R_4$ | $R_5$ | $R_6$ | $\mu$ | $\lambda_0$[Å] |
|---|---|---|---|---|---|---|---|---|
| $\bar{x}$ | 37.70 | 61.89 | 77.94 | 90.72 | 101.13 | 110.53 | 2.15052 | 8678.96 |
| $\sqrt{\sum_i (\bar{x} - x_i)^2 / N}$ | 0.41 | 0.21 | 0.16 | 0.15 | 0.15 | 0.25 | 0.00448 | 0.07969 |

Table 3

|  | $(O_{x1}, O_{y1})$ | $(O_{x2}, O_{y2})$ | $(O_{x3}, O_{y3})$ | $(O_{x4}, O_{y4})$ | $(O_{x5}, O_{y5})$ | $(O_{x6}, O_{y6})$ |
|---|---|---|---|---|---|---|
| $\bar{x}$ | (129.89, 130.73) | (129.95, 130.83) | (129.96, 130.92) | (129.99, 131.03) | (129.80, 130.99) | (129.69, 130.97) |
| $\sqrt{\sum_i (\bar{x} - x_i)^2 / N}$ | (0.19, 0.22) | (0.10, 0.09) | (0.08, 0.08) | (0.11, 0.10) | (0.14, 0.11) | (0.17, 0.16) |

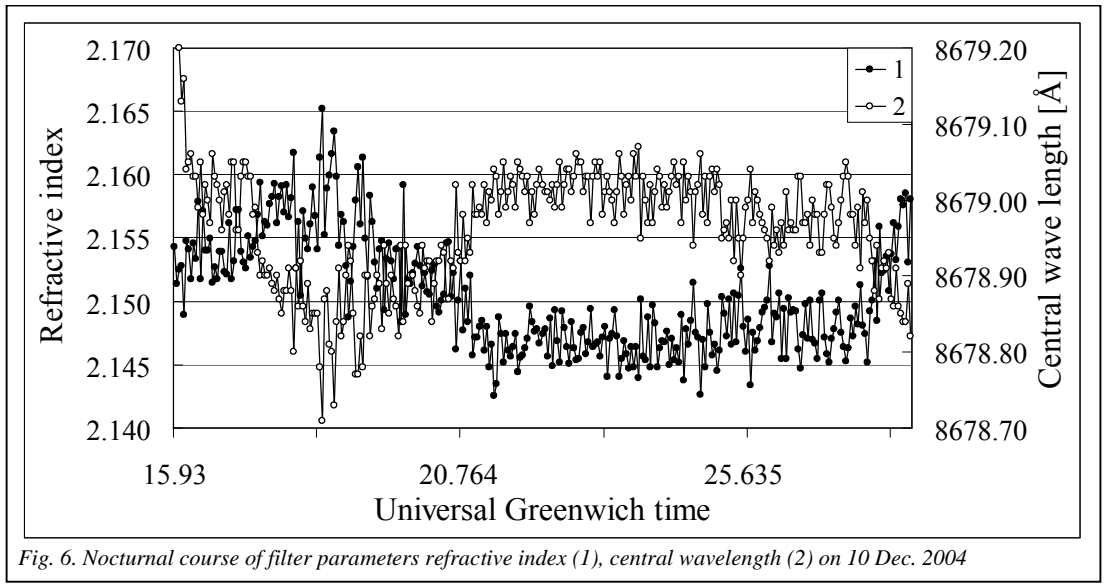

*Fig. 6. Nocturnal course of filter parameters refractive index (1), central wavelength (2) on 10 Dec. 2004*